# КОМП'ЮТЕРНІ ЗАСОБИ, МЕРЕЖІ ТА СИСТЕМИ

*O.V. Palagin, V.Yu. Velychko, K.S. Malakhov, O.S. Shchurov*

**PERSONAL RESEARCH INFORMATION SYSTEM. ABOUT DEVELOPING THE METHODS FOR SEARCHING PATENT ANALOGS OF INVENTION**

*The article describes information model and the method for searching patent analogs for Personal Research Information System.*

*В работе рассмотрены информационная модель АРМ и метод поиска аналогов изобретения среди патентной документации в произвольной предметной области.*
*Ключевые слова: автоматизированное рабочее место, композитный веб-сервис, атомарный веб-сервис, микросервис, патент.*

*В роботі розглянуто інформаційну модель АРМ та метод пошуку аналогів винаходу серед існуючої патентної документації у довільній предметній галузі.*
*Ключові слова: автоматизоване робоче місце, композитний веб-сервіс, атомарний веб-сервіс, мікросервіс, патент.*

О.В. ПАЛАГІН, В.Ю. ВЕЛИЧКО, К.С. МАЛАХОВ, О.С. ЩУРОВ

## АВТОМАТИЗОВАНЕ РОБОЧЕ МІСЦЕ НАУКОВОГО ДОСЛІДНИКА. ДО ПИТАННЯ РОЗРОБКИ МЕТОДІВ ПОШУКУ АНАЛОГІВ ПАТЕНТНОЇ ДОКУМЕНТАЦІЇ ВИНАХОДУ

**Вступ.** Створення засобів інформаційно-технологічної підтримки наукової діяльності людини, зокрема наукових досліджень, завжди було і є одним із центральних напрямків розвитку інформатики. Головними особливостями сучасного моменту розвитку наукових досліджень є трансдисциплінарний підхід і глибока інтелектуалізація всіх етапів життєвого циклу постановки і вирішення фундаментальних та прикладних наукових проблем. Крім завдань інфраструктурної підтримки наукових досліджень тут на перший план виходять завдання їхнього методологічного супроводу й забезпечення процесів інтеграції, конвергенції, уніфікованого представлення трансдисциплінарних знань і операцій над ними. Істотну роль, при цьому, відіграє системологічна підготовка навичок і розширення діапазону світогляду наукових дослідників з метою забезпечення двоєдності концепцій поглиблення знань у конкретній предметній області, з одного боку, і розширення проблеми, виходячи з реальності єдності світу й необхідності формування єдиної системи знань про світ, – з іншого.

Сучасні інформаційні системи підтримки наукових досліджень побудовані за принципами систем обробки даних та створення прикладних програмних систем для виконання окремих функцій (пошуку інформації, проектування різних виробів, проведення віртуальних експериментів, виконання математичних обчислень, тощо) або для деякого конкретної предметної області (підтримка проведення досліджень у фізиці, хімії, медицині, біології і т.п.). Наслідком цього є низька продуктивність професійної праці вчених і, в остаточному підсумку, недостатня ефективність виконання кожної із науково-технічних програм та відсутність синергетичного ефекту від галузевих наукових досліджень. Для підвищення ефективності науково-технічної діяльності за проектами "Програми інформатизації НАН України на 2015-2019 роки" [1] спроектований та програмно реалізований експериментальний макет комплексної знання-орієнтованої системи інформаційно-технологічної підтримки наукової діяльності – "Автоматизоване робоче місце наукового дослідника" (АРМ) [2]. АРМ побудована згідно загальних принципів розробки систем такого класу [3] (модульність, сервіс-орієнтована архітектура (а саме її варіант – мікросервісна архітектура [4,5]), кросплатформеність, використання хмарних обчислень (англ. *cloud computing, CC*), розробка програмного забезпечення за шаблоном Модель-Вигляд-Контролер (англ. *Model-View-Controller, MVC*), взаємодія між компонентами системи на основі JSON-моделі (англ. *JavaScript Object Notation, JSON*) даних та архітектурного стилю взаємодії компонентів розподіленого програмного забезпечення "передача репрезентативного стану" (англ. *Representational State Transfer, REST*), зберігання даних на основі абстракції сховища даних).

Важливою формою науково-технічної творчості є винахідницька діяльність. Інтелектуальна підтримка основних етапів життєвого циклу підготовки пакету заявочних документів на патентування винаходів в Україні, зокрема етапу пошуку аналогів винаходу серед існуючої патентної документації, є однією з центральних функцій АРМ. В основній частині роботи розглянемо узагальнену інформаційну модель АРМ, множину атомарних веб-сервісів АРМ та синтезовану ними множин функцій, розроблений метод пошуку аналогів винаходу серед існуючої патентної документації в довільній предметній галузі Міжнародної патентної класифікації (МПК) [6] та елементи його реалізації в рамках експериментального макету АРМ.

**Основна частина.** *Узагальнена інформаційна модель АРМ.*

Узагальнена інформаційна модель АРМ представляється у вигляді трикомпонентного кортежу композитного веб-сервісу (англ. *Composite Web Service, CWS*) з використанням формалізму наведеного в [7]:

$$CWS = \left\langle AWS, \{crd\}, F \right\rangle,$$

де:

$CWS$ – композитний веб-сервіс АРМ,

$AWS = \left\{ aws_i \mid i = \overline{1, m} \right\}_{m \in N}$ – множина атомарних веб-сервісів (англ. *Atomic Web Service, AWS*) доступних для використання,

$crd$ – компонент-координатор елементів $AWS$,

$F = AWS : \left\{ C_j \middle| j = \overline{1,n} \right\}_{n \in N}$ — множина функцій, кожна функція є результатом координування та взаємодії елементів $AWS$,

$C_j \subseteq AWS, C_j = \left\{ aws_k \middle| k \geq 1, k \leq m \right\}_{k \in N}$ — підмножина атомарних веб-сервісів необхідна для реалізації $j$-ої функції $CWS$.

**Множина атомарних веб-сервісів AWS APM.** На теперішньому етапі розробки програмного забезпечення АРМ множина $AWS$ складається з наступних атомарних веб-сервісів (які представляють собою мікросервіси (МкС), що дозволяє використовувати їх як самостійне програмне забезпечення відокремлене від АРМ так і як його компоненти) $aws$:

$aws_1$ – МкС екстракту тексту з оригіналів текстових документів формату pdf та його представлення у стандартному текстовому форматі plain/text (надалі тексту);

$aws_2$ – МкС екстракту тексту з оригіналів текстових документів формату doc/docx;

$aws_3$ – МкС визначення мови тексту;

$aws_4$ – МкС автоматичного реферування тексту для української мови;

$aws_5$ – МкС конвертації кодування тексту з UTF-8 у WIN-1251;

$aws_6$ – МкС автоматичного визначення назви, авторів та кількості сторінок для документу формату pdf;

$aws_7$ – МкС автоматичного визначення ключових слів тексту для української мови;

$aws_8$ – МкС автоматичного розбиття тексту на речення для української мови;

$aws_9$ – МкС автоматичного розбиття тексту на слова (англ. *Token*) – лематизовані словоформи та їх розмітку частинами мови (англ. *Part-of-speech tagging, PoS tagging*) для української мови;

$aws_{10}$ – МкС автоматичного розбиття тексту на слова – не лематизовані словоформи для української мови;

$aws_{11}$ – МкС автоматичного розбиття тексту на композиціональні терміни – композиціональна лінгвістична обробка тексту (англ. *Compositional Language Pre-processing, CLP*) – багатослівні й однослівні терміни, їх розмітку частинами мови, формування масиву з цих термінів для української мови;

$aws_{12}$ – МкС автоматичного видалення стоп-слів (слова, які не несуть смислового навантаження) з тексту;

$aws_{13}$ – МкС лематизації слів для української мови;

$aws_{14}$ – МкС автоматичного синтаксичного аналізу речень на основі

SyntaxNet/DRAGNN: Neural Models of Syntax [8,9];

$aws_{15}$ – МкС пошуку наукових публікацій у зовнішніх бібліографічних базах даних, зокрема, Google Scholar (інтелектуальний агент автоматизованого пошуку наукових публікацій [10]);

$aws_{16}$ – МкС індексації та розмітки текстів для повнотекстового пошуку;

$aws_{17}$ – МкС документо-орієнтованої системи керування базами даних MongoDB (зберігання оригіналів текстових документів та текстів, онтологічних структур, онтологій, опрацьованих текстів у вигляді JSON-документів, нейронних мовних векторних моделей (МВМ));

$aws_{18}$ – МкС онтологічної репрезентації текстів (графічний редактор маніпулювання онтологіями та онтологічними структурами – ІТ-платформа онтологічних інформаційно-аналітичних експертних систем ТОДОС);

$aws_{19}$ – МкС "КОНФОР" (вирішення задач виділення закономірностей, класифікації, генерації таксономії, діагностики та прогнозування);

$aws_{20}$ – МкС "КОНСПЕКТ" (графічний інтерфейс користувача (ГІК) для управління поверхневим синтактико-семантичним аналізом природно-мовних текстів, композиціональною лінгвістичною обробкою текстів, виділення термінів і контекстів та формування їх у класи);

$aws_{21}$ – МкС опрацювання МВМ колекцій відкритих даних (англ. *Open Data Collections, ODC*) – обчислення семантичної близькості однослівних та багатослівних термінів в рамках обраної МВМ, обчислення семантичних асоціатів для однослівних та багатослівних термінів в рамках обраної МВМ, обчислення центру лексичного кластера набору однослівних та багатослівних термінів в рамках обраної МВМ, обчислення семантичної близькості між двома наборами однослівних та багатослівних термінів в рамках обраної МВМ;

$aws_{22}$ – МкС ГІК персоніфікованої онтологічної бази знань публікацій наукового дослідника [3];

$aws_{23}$ – МкС генерації та заповнення шаблонів документів, що дозволяють формувати вхідний потік документів, який надходить від заявника стосовно заявок на об'єкти інтелектуальної власності;

$aws_{24}$ – додаткові зовнішні веб-сервіси, мікросервіси, програмне забезпечення;

Координування та взаємодію елементів *AWS* забезпечує компонент-координатор – *crd*, реалізований у вигляді зворотного проксі-сервера (reverse proxy) задач та загального ГІК веб-застосунку АРМ (вибір та керування режимами роботи і функціями АРМ). Зворотній проксі-сервер задач також використовується для балансування мережного навантаження між декількома серверами (нодами) та розподілення задач між відповідними RESTful веб-API (англ. *Web application programming interface, Web API*).

***Множина функцій F АРМ.*** Функціональне наповнення АРМ представляється наступним набором синтезованих з множини ***AWS*** функцій:

$$F = AWS : \{C_1, C_2, C_3, C_4, C_5, C_6, C_7\} \text{ ,де:}$$

$C_1$ – функція автоматизованої побудови прикладних онтологій в довільній предметній області [11,12]:

$$C_1 = \{aws_1 \quad aws_3, aws_5, aws_6, aws_8, aws_9, aws_{11}, aws_{12}, aws_{14}, aws_{16} \quad aws_{20}\}$$

$C_2$ – функція дослідного проектування наукових публікацій в довільній предметній області [3]:

$$C_2 = \{aws_1 \quad aws_9, aws_{11} \quad aws_{13}, aws_{15} \quad aws_{18}, aws_{22}, aws_{24}\}$$

$C_3$ – функція персоніфікованого онтологічного опрацювання наукових публікацій [3]:

$$C_3 = \{aws_1 \quad aws_9, aws_{11} \quad aws_{13}, aws_{15} \quad aws_{18}, aws_{22}\}$$

$C_4$ – функція автоматизованої композиціональної лінгвістичної обробки текстів для побудови MBM (дистрибутивної репрезентації текстів):

$$C_4 = \{aws_1 \quad aws_3, aws_5, aws_8 \quad aws_{12}, aws_{20}, aws_{24}\}$$

$C_5$ – функція дистрибутивно семантичного аналізу колекцій відкритих даних представлених у вигляді текстових документів в довільних предметних областях або в певній предметній області:

$$C_5 = \{aws_{21}, aws_{24}\}$$

$C_6$ – функція автоматичного синтаксичного аналізу речень:

$$C_6 = \{aws_1 \quad aws_3, aws_8, aws_{14}, aws_{20}, aws_{24}\}$$

$C_7$ – функція підготовки пакету заявочних документів на патентування винаходів в Україні:

$$C_7 = \{aws_1 \quad aws_3, aws_{21}, aws_{23}, aws_{24}\}$$

***Метод пошуку аналогів винаходу серед існуючої патентної документації та елементи їх реалізації в рамках АРМ.***

В рамках програми інформатизації НАН України на 2017 рік за проектом "Розробка елементів технології інформаційно-онтологічної підтримки науково-технічної творчості" Інституті кібернетики імені В.М. Глушкова НАН України розроблено методи пошуку аналогів винаходу серед існуючої патентної документації, на основі композиціонального дистрибутивно-семантичного аналізу відкритих даних, представлених у вигляді текстових документів форматів pdf, doc та docx, що має назву – метод персоніфікованого пошуку семантично близьких патентів.

Представимо розроблений метод у вигляді діаграми діяльності (англ. *activity diagram*) уніфікованої мови моделювання (англ. *Unified Modeling Language*, UML), яка відображає потік робіт (англ. *Workflow*) та систематизовану сукупність дій, котрі вирішують задачу пошуку аналогів винаходу серед існуючої патентної документації в довільній предметній галузі МПК.

Діаграма діяльності методу персоніфікованого пошуку семантично близьких патентів представлена на рис. 1. Одними з центральних серед етапів та дій потоку робіт зазначеного методу слід особливо підкреслити:

- етап композиціональної лінгвістичної обробки завантаженої патентної документації (на цьому етапі документація представлена у вигляді plain/text). Цей етап є підготовчим кроком перед формуванням так званого "Нормалізованого лінгвістичного корпусу текстів" (НКТ) на основі якого навчається нейронна МВМ. Він забезпечує якість НКТ, що в свою чергу впливає на якість навченої нейронної МВМ і дозволяє отримати більш високу точність при обчисленні косинусної близькості патентної документації. Композиціональна лінгвістична обробка патентної документації включає: розбиття тексту патенту на речення; екстракт термінів з тексту патенту (в тому числі формування композиціональних термінів), розмітка частинами мови лексем та композицій лексем з тексту патенту, вилучення стоп-слів або шумових слів з тексту патенту (слів, які не несуть смислового навантаження в тексті, зокрема прийменники, суфікси, дієприкметники, вигуки, цифри тощо), індексація та розмітка тексту патенту для повнотекстового пошуку та збереження онтологічних даних та знань в персоніфікованій базі даних АРМ;
- етап композиціональної лінгвістичної обробки нової патентної документації. Цей етап є підготовчим кроком нового патенту до порівняння з існуючими патентами в рамках нейронної МВМ та включає екстракт термінів з тексту патенту (в тому числі формування композиціональних термінів), розмітка частинами мови лексем та композицій лексем з тексту патенту, індексація та розмітка тексту патенту для повнотекстового пошуку та збереження онтологічних даних та знань в персоніфікованій базі даних АРМ;
- етап навчання нейронної МВМ. На цьому етапі з використанням вільного програмного інструментарію дистрибутивного аналізу текстів: word2vec [13], fasttext [14] та програмної бібліотеки для досліджень в області обробки природньої мови (англ. *Natural language processing, NLP*) та векторного моделювання (англ. *Vector space model*) gensim [15], яка включає API для роботи з алгоритмами word2vec, fasttext та інші, виконується навчання нейронної МВМ;
- етап ініціалізації навченої нейронної МВМ. На цьому етапі навчена нейронна МВМ проходить процедуру ініціалізації на окремому сервері зі спеціалізованим програмним забезпеченням (включаючи програмну бібліотеку gensim), яке адаптоване для використання архітектурного стилю REST;
- етап обчислення коефіцієнту косинусної близькості (семантичної схожості) між масивами термінів нового та існуючих патентів.

Обчислення здійснюється використовуючи API програмної бібліотеки genism, а саме, функція n_similarity(ws1, ws2) з пакету models.keyedvectors.

На основі обчислення коефіцієнту косинусної близькості (семантичної схожості) формується фінальний список патентів, які є найближчими аналогами до нового патенту. Коефіцієнт косинусної близькості (семантичної схожості) патентів може приймати значення в проміжку [-1 ... 1]. Якщо коефіцієнт косинусної близькості (семантичної схожості) патентів приймає значення в проміжку [-1 ... 0,5] – це свідчить про відсутність схожих контекстів (англ. *word embeddings*) та найменшу семантичну близькість патентів. Якщо коефіцієнт косинусної близькості (семантичної схожості) патентів приймає значення в проміжку [0,5 ... 1] – це свідчить про наявність схожих контекстів та більшу семантичну близькість патентів. Чим більше коефіцієнт косинусної близькості (семантичної схожості) наближається до 1, тим більша семантична близькість документів та більше схожих контекстів в патентах. НТК, Нейронна МВМ та нова патентна документація є персоніфікованою для кожного користувача АРМ.

**Висновки.**

В роботі розглянуто узагальнену інформаційну модель АРМ, множину атомарних веб-сервісів АРМ та синтезованих ними множин функцій, метод пошуку аналогів винаходу серед існуючої патентної документації в довільній предметній галузі МПК та елементи його реалізації на основі композиціонального дистрибутивно-семантичного аналізу відкритих даних в рамках експериментального макету АРМ. Розроблений експериментальний макет АРМ (у вигляді композитного веб-сервісу) значно підвищує ефективність роботи наукового дослідника в такий важливій формі науково-технічної творчості, як винахідницька діяльність, зокрема підготовку пакету заявочних документів на патентування винаходів в Україні. В подальшій роботі планується провести дослідження в області автоматизованого синтезу композитних веб-сервісів в залежності від поставленої задачі у вигляді алгоритму її виконання та вирішення, а також розробити та впровадити на основі АРМ комплексний підхід до автоматизованого синтезу композитних веб-сервісів.

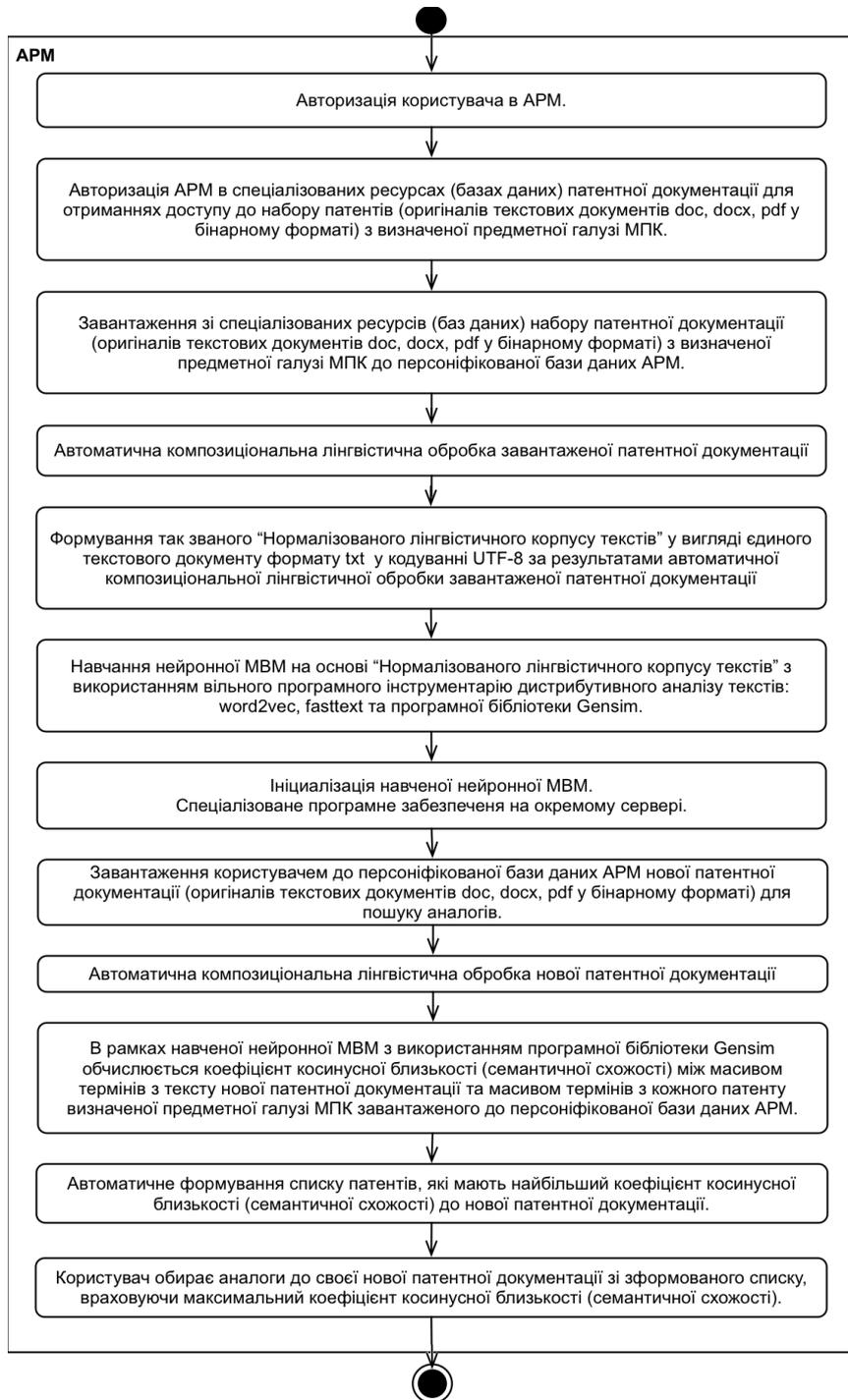

РИС. 1. Діаграма діяльності методу персоніфікованого пошуку семантично близьких патентів

## Література